\documentclass[12pt]{article}
\usepackage{graphicx}
\begin{document}
\begin{center}
{\bf General Relativistic Thermoelectric Effects\\ in
Superconductors}\\

\vspace*{1cm}

{\bf B.J. Ahmedov}\footnote{E-mail: ahmedov@astrin.uzsci.net}\\

\vspace{1cm} {\em Institute of Nuclear Physics and Ulugh Beg
Astronomical Institute\\ Ulughbek, Tashkent 702132, Uzbekistan}\\
\end{center}
\vspace{2cm}
\begin{abstract}

We discuss the general-relativistic contributions to occur in the
electromagnetic properties of a superconductor with a heat flow.
The appearance of general-relativistic contribution to the
magnetic flux through
 a superconducting thermoelectric bimetallic circuit is shown.
A response of the Josephson junctions to a heat flow is
investigated in the general-relativistic framework. Some
gravitothermoelectric  effects which are observable in the
superconducting state in the Earth's gravitational field are
considered.

\end{abstract}

\newpage

\section{Introduction}

Effects of fields of gravity and inertia on a superconductor have
been investigated by a number of authors starting DeWitt [1] and
Papini [2]. More recently, a general relativistic treatment for
electromagnetic effects in normal conductors with a heat flow and
superconductors without gradient of temperature has been given. In
particular, several superconducting devices that can, in
principle, detect a gravitational field have been presented by
Anandan [3]. Further, one of them has been already realized in the
experiment [4] which tested the equivalence principle for Cooper
pairs.

From our point of view a study of thermoelectromagnetic relativistic
gravitational effects in superconductors with nonzero gradient of
temperature is
of fundamental interest due to the following two reasons.

Thermoelectric effects do not vanish, in principle, in
inhomogeneous and anisotropic superconductors and nowdays attract
an increasing interest [5] especially due to the discovery of high
temperature superconducting materials which combine both
anisotropy and inhomogeneity properties and which, in this
connection, are favourable for measurement of thermal effects.
Since superconductors provide extremely sensitive and accurate
measurements, there is  hope that the very weak general
relativistic contributions to the thermoelectromagnetic effects
due to the inhomogeneity arising from the Earth's gravitational
field might be detectable.

On the other hand, according to the recent theoretical models,
core of neutron stars forms matter in a superconducting state with
thermal distribution and heat flow (see, for example, [6]). It
suggests that thermoelectromagnetic effects might explain of
origin and evolution of magnetic field inside the core of
supermassive pulsars for which the dimensionless general
relativistic compactness parameter ${\alpha}/{R_s}$ reaches about
$0.5$ [7] ($\alpha$ and $R_s$ are the gravitational radius and
radius of star, respectively). In addition thermoelectromagnetic
effects in Josephson contacts could be responsible for
electromagnetic radiation arising from possible
superconductor-normal metal-superconductor layers in the
intermediate boundary between conducting crust and superconducting
core inside neutron star.

\section
{General Relativistic Correction to the Magnetic Flux Induced by a
Heat Flow in Superconducting Thermocouple}

Consider a superconductor with a heat flow in the external
stationary gravitational
field. According to the two-component model two currents flow in this
superconductor: the superconducting
current of density $\hat\jmath_{(s)\alpha}$ and the normal current of
density $\hat\jmath_{(n)\alpha}$. The normal current is carried by
`normal' electrons (exitations) and it does not differ essentially from
the current in the normal state of a metal [8]
\begin{eqnarray}
F_{\alpha\beta}u^{\beta}=\frac{1}{\lambda}\hat\jmath _{(n)\alpha}+
R_H(F_{\nu\alpha}+u_\alpha u^\sigma F_{\nu\sigma})\hat\jmath_{(n)}^\nu
+\Lambda^{-1/2}\stackrel{\perp}{\nabla}_\alpha\tilde\mu -
\beta\Lambda^{-1/2}\stackrel{\perp}{\nabla}_\alpha\tilde
T-b\hat\jmath_{(n)}^\beta A_{\alpha\beta},
\end{eqnarray}
where $\tilde T=\Lambda^{1/2}T, \tilde\mu =\Lambda^{1/2}\mu$, $\mu$ is
the chemical potential per unit charge, $T$ is the temperature, $\lambda$
is the
conductivity, $\Lambda =-\xi^\alpha\xi_\alpha$, $\xi_\alpha$ is a timelike
Killing vector being parallel to the four velocity of the conductor
$u^\alpha$, $\beta$ is normal differential thermopower coefficient,
$F_{\alpha\beta}=A_{\beta ,\alpha}-A_{\alpha ,\beta}$ is
the electromagnetic
field tensor, $A_\alpha$ is the vector potential, $R_H$ is the Hall
constant,
$A_{\beta\alpha}=u_{[\alpha ,\beta]}+u_{[\beta}w_{\alpha]}$ is the
relativistic rate of rotation, $w_\alpha=u_{\alpha;\beta}u^\beta$ is the
absolute acceleration and $b$ is the
parameter for the conductor, $\stackrel{\perp}{\nabla}_\alpha$ and
$[\cdots]$ denote the
spatial part of covariant derivative and antisymmetrization.

Thus the electric current flowing in the conductor in the general case arises
for the following reasons (with general relativistic corrections and
contributions) - (a) electric field, (b) Hall effect, (c) nonequilibrium
effects and (d) Coriolis (gravitomagnetic) force effects described by the
last term on the right hand side of equation (1).

If the wavefunction of the Cooper pairs is $\psi=n_s^{1/2}
e^{i\vartheta}$ then four-vector of supercurrent density is [3]
\begin{eqnarray}
j_{(s)\alpha}=\frac{ie\hbar}{m_s}\{
\psi^*(\partial_\alpha-i\frac{2e}{\hbar c}A_\alpha)\psi-
\psi (\partial_\alpha+i\frac{2e}{\hbar c}A_\alpha)\psi ^*\},
\end{eqnarray}
and satisfies to equation
\begin{eqnarray}
j_{(s)\alpha}=\frac{2n_se}{m_s}
\{-\hbar\partial _\alpha\vartheta +2e/cA_\alpha\}=
\frac{2n_se}{m_s}P_\alpha
\end{eqnarray}
where
$n_s^{1/2}$ represents the density of Cooper pairs, $m_s$ and $P_\alpha=
p_\alpha+\frac{2e}{c}A_\alpha$
are the mass and the generalized momentum of
the Cooper pair and $\vartheta$ is the phase of superconducting
wavefunction.

Since any supercurrent flows only on the surface within the penetration
depth, in the interior of the superconductor supercurrent (3)
is parallel to the four-velocity $u^\alpha$ i.e.
$P^\alpha=-\frac{2e}{c}\mu
u^\alpha$. It follows from it that interior of the superconductor
\begin{eqnarray}
F_{\alpha\beta}={2}\mu_{[,\beta} u_{\alpha
]}+{2}\mu\partial_{[\beta}u_{\alpha]}\nonumber
\end{eqnarray}
and after multiplying it by $u^\alpha$ one can get that [3]
\begin{eqnarray}
E_\beta-\Lambda^{-1/2}\partial_\beta\tilde\mu=0
\end{eqnarray}
everywhere inside the superconductor in the steady state.

Formula (4) can be also derived from the general relativistic London
equations [9]
\begin{eqnarray}
\hat\jmath_{(s)[\beta;\alpha]}-\hat\jmath_{(s)[\alpha}(\ln n_s)_{,\beta]}
+\frac{2n_se^2}{m_sc}F_{\alpha\beta}-
{2cen_s}(A_{\alpha\beta}+u_{[\alpha}w_{\beta]})=0,
\end{eqnarray}
which have been obtained by requiring,
that inside a superconducting medium, the motion of the Cooper pairs
is free of resistance.

Suppose that two ends of a bulk piece of superconductor are at different
temperatures, $T_1$ and $T_2$. The temperature gradient will produce a
force on normal exitations of the superconductor initiating a current of
the normal exitations
\begin{eqnarray}
\hat\jmath_{(n)\alpha}=\Lambda^{-1/2}\lambda\beta\partial_\alpha\tilde T
-\lambda R_H(F_{\nu\alpha}+u_\alpha u^\sigma
F_{\nu\sigma})\hat\jmath_{(n)}^\nu
+\lambda b\hat\jmath_{(n)}^\beta A_{\alpha\beta}
\end{eqnarray}
as a consequence of Ohm's law (1) under the conditions corresponding to
equation (4).

It therefore follows that below $T_c$ under steady-state conditions
when equation (4) should be obeyed, if $\partial_\nu\tilde T\ne0$, the
density of the normal current $\hat\jmath_{(s)\alpha}$ should be finite
because of equation (6). However, if the circuit is open, then the total
current is zero and in the simplest case the density of the total current
also vanishes: $\hat\jmath_{\alpha}=\hat\jmath_{(s)\alpha}+\hat\jmath_{(n)
\alpha}=0$, i.e. the normal current density $\hat\jmath_{(n)}$ is
cancelled locally by a counterflow of supercurrent density
$\hat\jmath_{(s)}$
\begin{eqnarray}
\hat\jmath_{(s)\alpha}=-\hat\jmath_{(n)\alpha}=
-\Lambda^{-1/2}\lambda\beta\partial_\alpha\tilde T
+\lambda R_H(F_{\nu\alpha}+u_\alpha u^\sigma
F_{\nu\sigma})\hat\jmath_{(n)}^\nu
-\lambda b\hat\jmath_{(n)}^\beta A_{\alpha\beta}.
\end{eqnarray}

Suppose the superconductor is embedded in the Schwarzschild space-time
\begin{eqnarray}
ds^2=-(1-\alpha /r)(dx\circ )^2+(1-\alpha /r)^{-1}dr^2+r^2(d\theta^2+
\sin^2\theta d\varphi^2),
\end{eqnarray}
where timelike Killing vector
$\xi^\alpha$ can be chosen so that $\Lambda ={1-\alpha /r}$.
If curvature effects are negligible, than the apparatus may be regarded as
having an acceleration, $g$, relative to a local inertial frame, and thus,
$\Lambda=(1+2gH/c^2)^2$, where $H$ is the height above some fixed point.

Because of the cancellation of the thermoelectric current in the
superconductor, schemes for measuring the thermoelectric effects, based on
inhomogeneous or anisotropic superconductor configurations, become
necessary. A review of various experiments, as well as a bimetallic
superconducting ring has been given e.g. by Van Harlingen [10].

\begin{figure}
\begin{center} \label{fig1}
\includegraphics[height=60mm, width=86mm]{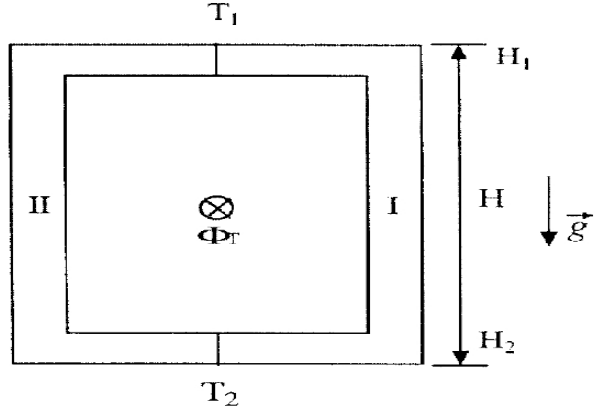}
\end{center}
\caption{Totally superconducting circuit made of two metals $I$
and $II$ in the Earth's vertical gravitational field. The magnetic
field $\Phi_T$ through the ring has gravitational contribution
proportional to $gH/c^2$.}
\end{figure}

We shall now consider two bulk specimens of two dissimilar
superconductors, $S_{I}$ and $S_{II}$, which are brought into
contact in such a way that together they form a closed ring, as in
Figure 1. Suppose that the temperatures of the upper and lower
contacts are kept at the different temperatures $T_1$ and $T_2$
respectively. The presence of an open gap in a massive circuit
when its thickness $d$ (for example, the diameter of the wire
forming the circuit)  is much bigger than the depth of penetration
$\delta$ of the field allows us to calculate the magnetic flux
$\Phi_b$ across the gap without solving the problem completely. As
it was shown above in the bulk of a superconductor
$\hat\jmath_\alpha=\hat\jmath_{(s)\alpha}+\hat\jmath_{(n)\alpha}=0$
and so
\begin{eqnarray}
\hat\jmath_{(n)\nu}=-\Lambda^{-1/2}\lambda\beta\partial_\nu\tilde
T=-\hat\jmath_{(s)\nu}=
\frac{2n_se}{m_s}[\hbar\partial_\nu\vartheta-\frac{2e}{c}A_\nu].
\end{eqnarray}
For the sample embedded in the Schwarzschild space-time last two terms in
formula (7) disappear
since in this case magnetic field does not penetrate inside the
superconductor and all components the relativistic rate of rotation are zero.

Integrating the equation (9) along the contour $C$, which is
in the bulk of the superconductor and using that $\oint A_\alpha
dx^\alpha= \frac{1}{2}\int F_{\alpha\beta}dS^{\alpha\beta}=\Phi_b$ and
$\oint\partial_\nu\vartheta dx^\nu=2\pi n$, where $n=0,1,2,...$ yields
directly
\begin{eqnarray} \Phi_b=n\Phi_0+ \frac{m_sc}{4n_se^2}
\int\Lambda^{-1/2}\lambda\beta\partial_\nu \tilde T, \end{eqnarray}
where
$m_s/e^2n_s=\Lambda_0=4\pi\delta^2/c^2$ and
$\Phi_0=\pi\hbar c/e=2\times 10^{-7}Gauss\cdot
cm^2$ is quantum of the magnetic flux. The current $I_s$ which leads to
the appearance of a flux $\Phi_b$ flows on the internal surface of the
circuit in a layer of thickness of the order of $\delta$.

If $\Lambda=\Lambda (0)$ at $0$ then from (10),
the magnetic flux through contour is
\begin{eqnarray}
\Phi_b=n\Phi_0+\Lambda (0)^{-1/2}\frac{m_sc}{4n_se^2}
\oint\beta\lambda\partial_\nu \tilde Tdx^\nu=\nonumber\\
n\Phi_0+\frac{m_sc}{4n_se^2}((\beta\lambda)_I-(\beta\lambda)_{II})
[T_1(1+gH_1/c^2)-T_2(1+gH_2/c^2)],
\end{eqnarray}
where $\beta_I$ and $\beta_{II}$ are the values of $\beta$ for the
two metals, $H_1$ and $H_2$ are the heights of the junctions above
the Earth's surface. This is the general relativistic
generalization of the thermoelectric effects in the inhomogeneous
(bimetallic) superconductor.

If, for the sake of simplicity, we assume that $(\beta\lambda)_{II}\gg
(\beta\lambda)_{I}$  we
then find from equation (11) that
\begin{eqnarray}
\Phi_T=\Phi_b -n\Phi_0\approx\frac{m_sc}{4n_se^2}(\beta\lambda
)_{II}\Delta T+\frac{gH}{c^2}\frac{m_sc}{4n_se^2}(\beta\lambda
)_{II}\Delta T.
\end{eqnarray}

When apparatus is horisontal, the magnetic flux in the circuit would remain
flat space-time value and last term in formula (12) will disappear.
In this case equation (12) describes unquantized thermoelectric flux
$\Phi_T$ in the presence of a temperature gradient. For typical parameters
of the superconductor the thermal-current-related flux (first term in
(12)) is expected to be of order of $10^{-2}\Phi_0$. In actual fact,
however, much - indeed orders of magnitude - stronger fluxes of hundreds
of $\Phi_0$ have been measured experimentally [11].

However if the apparatus is brought into a vertical plane, then
the magnetic flux will be changed according to formula for
$\Phi_T$. The height between two junctions is changed by
$H=H_1-H_2$ and the magnetic flux undergoes a fractional change
$gH/c^2$. Then the general relativistic contribution in $\Phi$
when $H=10cm$, in the Earth's gravitational field, will be
proportional to small dimensionless parameter $10^{-17}$.

There is no doubt the flux production mechanism discussed in this section
may be significant and relevant to the problem of origin and evolution of
magnetic fields in isolated neutron stars since their radius can be only
$1.4-3.5$ times larger than gravitational radius $\alpha$ and their
substance may be superconducting of $II$-type or superfluid at high
densities (see e.g. [12]).

\section{Thermoelectric Effects in $SNS$ Junctions in the External
Gravitational Field}

In this section we shall first consider the behaviour of an
superconductor-normal metal-superconductor ($SNS$) junction when its $S$
electrodes have different temperatures or, in other words, when there is a
heat flow through the junction placed in a gravitational field. After
that, we shall discuss the thermoelectric effects arising when, in
addition to carrying a heat flow, the junction is placed in both magnetic
and gravitational fields.

Let us suppose that there is a temperature difference $\Delta T$
between superconducting electrodes of the $SNS$ Josephson junction
placed in the external gravitational field (8) (see Figure 2). Due
to the
 Kirchoff's first law
\begin{eqnarray}
\hat\jmath _{(s)\alpha}=\hat\jmath _
{(n)\alpha}.
\end{eqnarray}

\begin{figure}
\begin{center} \label{fig2}
\includegraphics[height=60mm, width=86mm]{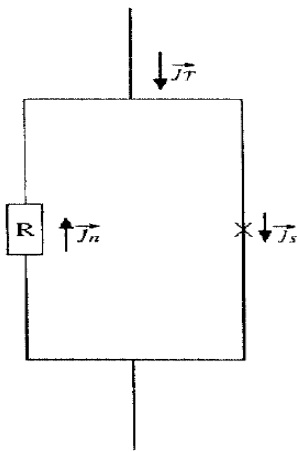}
\end{center}
\caption{Equivalent circuit for $SNS$ junction with heat flow:
$\vec\jmath_T=\lambda\beta\Lambda^{-1/2}grad\tilde T$ is the
current due to the heat flow across the normal layer.}
\end{figure}

 According to Ohm's law (1) a normal component of a current
in the junction is
\begin{eqnarray}
\hat\jmath _{(n)\alpha}=\lambda E_\alpha
-\lambda\Lambda^{-1/2}\partial_\alpha\tilde\mu +
\beta\lambda\Lambda^{-1/2}\partial_\alpha\tilde T.
\end{eqnarray}

Density of the superconducting current flowing through the junction is
related to the phase difference $\phi=\Delta\vartheta$ across the junction
by
\begin{eqnarray}
\hat\jmath _{(s)\alpha}=\hat\jmath _{(c)\alpha}\sin\phi
\end{eqnarray}
where $\hat\jmath_{(c)\alpha}$ is the critical value of electric
current density.

Using formulae (13)-(15) and Josephson equation [3]
\begin{eqnarray}
-\hbar\frac{\partial\phi}{\partial\tau}=\frac{2e}{c}\mu-\frac{2e}{c}
A_\mu\xi^\mu\Lambda^{-1/2}
\end{eqnarray}
one can obtain
\begin{eqnarray}
\hat\jmath_{(c)\alpha}\sin\phi =\lambda\Lambda^{-1/2}\{-\partial_\alpha
(\frac{\hbar c}{2e}\frac{\partial\phi}{\partial\tau}\Lambda^{1/2})+
\beta\partial_\alpha\tilde T\}\nonumber
\end{eqnarray}
which after integration on $n^\alpha dS$ will give
\begin{eqnarray}
\Lambda (0)^{1/2}RI_{(c)}\sin\phi
=-\frac{\hbar c}{2e}\frac{\partial\phi}{\partial\tau}\Lambda^{1/2}+
\beta\Delta\tilde T.
\end{eqnarray}
Here $R=\int\frac{dl}{\rho dS}$ is the resistance of the normal layer with
length $dl$,
$I_{(c)}=\int \hat\jmath_{(c)\alpha}n^\alpha dS$ is electric current,
$n^\alpha$ is normal vector to the cross section of wire $dS$,
$\partial\phi /\partial\tau=\phi_{,\alpha}u^\alpha$.

Therefore if thermoelectric current exceeds the critical
current of the Josephson junction, then as a consequence of (15) and (17),
an alternating current (ac) of frequency
\begin{eqnarray}
\omega=\frac{2e}{\hbar}\beta\Lambda^{-1/2}\Delta\tilde T
\end{eqnarray}
is produced and the junction emits radiation with the frequency $\omega$
measured by an observer at rest with respect to the junction.
Formula (18) is the general relativistic generalization of thermoelectric ac
Josephson effect according to which a temperature difference $\Delta T$
across the $SNS$ junction results in electromagnetic radiation, which has
been predicted in [13] and experimentally confirmed [14].

Now we would like to show with help of simple example
that general relativistic thermoelectric effects may have astrophysical
applications.
Suppose that $SNS$ junctions can be realized in the intermediate boundary
between conducting crust and superconducting core inside neutron star.
Then as a consequence of equation (18), that is,  of the existence of
thermal analog of the ac Josephson effect, we can predict a
new mechanism for electromagnetic radiation production from pulsars:
 a steady heat flow through SNS junctions in intermediate boundary gives
rise to Josephson radiation with frequency $\omega$. Let us
roughly estimate the frequency of expecting radiation. If
thermoelectric power of conducting crust $\beta (T\sim 10^{5\circ
}K)\sim 10^{-2}V/   K$, then $\omega\approx 3\times 10^{13}
\Lambda^{-1/2}(T_2-T_1)   K$, and at a temperature difference
$10^{-7 \circ}K$ we have $\omega\sim 3\times 10^5 Hz$.

According to (18) the frequency of the junction depends from
altitude in gravitational field and hence frequences $\omega_1$
and $\omega_2$ of the junction at $H_1$ and $H_2$ ($H=H_2-H_1;
H_1<H_2$) are connected through $\omega (H_2)=\omega
(H_1)(1-gH/c^2)$. If $\beta \approx 10^{-5}V\cdot   K^{-1}$ and
$H=10m$, then the general relativistic red-shift for the frequency
is
\begin{eqnarray}
\omega_{gr}(Hz)=\frac{gH}{c^2}\frac{2e}{\hbar}\beta \Delta
T\approx 3\times 10^{-3}\Delta T(  K).
\end{eqnarray}
This quantity is weak since it is experimentally very difficult to obtain
a great temperature difference at the superconducting system. If the
radiations are detected by one device, there would be no frequency
difference between them due to the gravitational redshift effect for
frequency which will compensate correction (19). Hence this null
experiment would confirm that the temperature difference, $\Delta
T_2$ and $\Delta T_1$, at the upper and low altitudes are connected
through
\begin{eqnarray}
\Delta T_1=\Delta T_2(1-gH/c^2)
\end{eqnarray}
and varying with height.

Consider two dissimilar SNS junctions, separated by a height $H$ and
connected in
parallel by superconducting wires to a common heater source.
Suppose that $\beta _{II}$ and $\beta _{I}$ are the thermoelectric power
of the upper and lower junctions, respectively. Now we integrate equation
(3) over the contour which passes through the interior of superconducting
ring with two Josephson contacts. Then
\begin{eqnarray}
\Phi_b=n\Phi_0+\frac{\hbar c}{2e}(\phi_{II}-\phi_{I}),
\nonumber
\end{eqnarray}
where $\phi_{II}$ and $\phi_{I}$ are the contributions due to the phase
discontinuities at the Josephson junctions.

The rate of change of magnetic flux is related to $\Delta\omega
\equiv \omega_{II}-\omega_I$ by
\begin{eqnarray}
\frac{d\Phi_b}{d\tau}=\frac{dn}{d\tau}\Phi_0+\frac{\hbar c}{2e}\Delta\omega ,
\nonumber
\end{eqnarray}
so if $\Delta\omega$ is not equal to zero, a magnetic flux
$\Delta\Phi_b\ne 0$ will be induced. As long as $\Delta\Phi_b<\Phi_0$,
$n$ will remain constant and $\Delta\Phi_b$ will increase linearly with
time until $\Delta\Phi_b=\Phi_0$, then the order of the step $n$ will
change as flux quantum enters the loop. Thus due to the effect of
gradient of temperature on the junctions with unequal
thermoelectric powers is equivalent to having a time dependent flux, as
given by the last term on the right hand side of the last equation.
 Then using equations (18) and (20) one can derive that the change in
magnetic field inside the circuit during the time interval $[0,\tau]$ is
\begin{eqnarray}
\Delta\Phi_b ={c}\int_{0}^{\tau}[\Delta T_1\beta_{I}-\Delta
T_2\beta_{II}(1-\frac{gH}{c^2})]d\tau +\Delta n\Phi_0=\nonumber\\
{c}\int_{0}^{\tau}\Delta T_1 (\beta_{I}-
\beta_{II})d\tau +\Delta n\Phi_0.
\end{eqnarray}

Thus this particular loop is sensitive to the frequency and in this
connection to the thermoelectric power difference between the junctions.
 The independence of magnetic flux (21) from gravitational field $g$
confirms the validity of formula (18) for the temperature, that is
in a gravitational field, during thermal equilibrium, the related
quantity $\tilde T$ (rather T) is constant along the sample. In
addition a new experiment, in which the thermoelectric response
creates flux (21) changing with time will give one more
possibility of measurement of thermoelectric effects in
superconductors. It is more or less important, not only from
general-relativistic point of view, but also for the new proposals
[15] for confirmation of some aspects of thermoelectric transport
theory.

When the current exceeds the critical value potential difference
$\tilde V=\beta\Delta\tilde T$ appears across the junction due to
the thermoelectric effects. Since the thermoelectric power of the
junction $II$ differs from that at the junction $I$, the potential
differences across the first and second junctions, $\tilde V_{II}$
and $\tilde V_I$, respectively, will differ so that $\Delta\tilde
V=\tilde V_{II}-\tilde V_I=(\beta_{II}-\beta_{I})\Delta \tilde T.$
The basic technique for the detection of extremely small voltage
differences between two Josephson junctions by monitoring of
magnetic flux change was firstly developed by Clarke [16].

In the absence of any additional effects on the Cooper pairs, one
would thus expect the net EMF in the loop containing the junctions
to be $(\beta_{II}-\beta_{I})\Delta\tilde T\sim 10^{-11} V$ for
the typical values of parameters $(\beta_{II}-\beta_{I})\sim
10^{-6} V/  K$ and $\Delta\tilde T\sim 10^{-5\circ}K$. For the
loop of inductance $L$ the evolution of magnetic field is
approximately governed by law
$\frac{d\Phi_b}{d\tau}=-L\frac{dI_l}{d\tau}$. In this connection a
nonvanishing value for $\Delta V$ would lead, according to (18)
and (21), to a time varying current $I_l$ (from zero to the
critical maximum value in the range of one number of the step
$n$): $\frac{dI_l}{cd\tau}=-\frac{1}{L}\Delta V$, which will
induce the above discussed flux $\Delta\Phi_b=c\int\Delta Vd\tau$
through the loop in the linear regime.

For measurement of gravitational contribution we can propose the following
change. Connect the Josephson junctions to two independent heaters which
have different temperatures, such that $\Delta T_2=A\Delta T_1$ and $A$ is
constant. In this case the flux (21) takes form
\begin{eqnarray}
\Delta\Phi_b =\Delta n\Phi_0+
{c}\int_{0}^{\tau}\Delta T_1 (\beta_{I}-A\beta_{II})d\tau
+A{c}\int_{0}^{\tau}\Delta T_1\beta_{II}\frac{gH}{c^2}d\tau .
\end{eqnarray}
If the apparatus is horisontal then $H=0$ and therefore the last
term in the magnetic flux change should disappear. When the coil
is vertical the last term will increase the rate of change of flux
according to (22). By detecting this contribution, the
gravitational corrections to thermoelectric effects can be
measured. Taking $A\beta\sim 10^{-7}cm^{1/2}\cdot g^{1/2}\cdot
s^{-1}/  K, \Delta T_1\sim 10^{-4\circ} K$ and $H=10cm$ we obtain
for $\Delta\Phi_G(Gauss\cdot cm^2)\sim 3\times 10^{-19} \cdot
\Delta\tau (s)$. Measuring such tiny variations of magnetic field
for large $\Delta\tau$ is near to
 the limit of SQUID sensitivity.

The main problem in observation in the flux (21) and (22) will be connected
with generating thermal current $\vec\jmath_s=-\Lambda^{-1/2}\lambda\beta
grad\tilde T$ comparable in magnitude with its critical value $\vec\jmath_c$,
since the temperature difference across the junction is limited by a low
temperature $T_c$ and small sizes of the junction.

The similar method of measurement has been used by Jain et al [4] in null
result experiment on confirmation of the strong equivalence principle for a
charged massive particle. In their experiment the
phase of Josephson contacts has been locked to an external microwave source
and is schown to be technically feasible to measure a voltage drop $10^{-22}V$.

It is interesting to mention that the predicted mechanism for
production of magnetic field and current changing with time can be
of crucial importance in astrophysics as a way (additional to the
existed ones [17]) for generation of electromagnetic radiation
from pulsars. According to the recent theoretical models [7], a
neutron star is the relativistic compact object consisting of the
conducting crust and superfluid core. In the inner crust of the
neutron star the superfluid coexists with a crystal lattice and in
its core, at densities above $2\times 10^{14} gm/cm^3$ there is a
homogeneous mixture of superfluid neutrons and superconducting
protons.

Important fact is that the thermoelectric power $\beta$ is the function of
temperature as $T^{3/2}$ and in this connection can reach large numbers
since superconductivity in
the stars takes place at the temperatures $10^6\div 10^{7\circ}K$. So if
one accepts that the $SNS$ structures are realized in the intermediate
boundary between conducting crust and superconducting core inside the
neutron star then the strong heat fluxes in these $SNS$ junctions can
lead to the generation of time-varying magnetic field (i.e. electromagnetic
radiation) due to the thermoelectric effect described by the basic formula
(21).

Suppose that, in addition to the heat flow, $SNS$ Josephson junction is
placed in a magnetic field parallel to the plane of the junction (the xy
plane). It is well-known that the maximal current density,
$\hat\jmath_{(s)max\alpha}$, which can pass through the junction is
\begin{eqnarray}
\hat\jmath_{(s)max\alpha}=\hat\jmath_{(c)\alpha}|\frac{\sin\pi\Phi_b/\Phi_0}
{\pi\Phi_b/\Phi_0}|,
\end{eqnarray}
where $\Phi_b$ is the magnetic flux through the junction.

Taking into account that $\hat\jmath_{(s)\alpha}=
\lambda\beta\Lambda^{-1/2}\partial_\alpha \tilde T$ we can find
that the critical temperature difference, corresponding to appearance of
voltage across the junction, is
\begin{eqnarray}
(\Delta\tilde T)_{c}=\Lambda (0)^{1/2}\frac{I_c
R}{\beta}|\frac{\sin\pi\Phi_b/\Phi_0}{\pi\Phi_b/\Phi_0}|
\end{eqnarray}
and depends from gravitational field, where $\Lambda =\Lambda (0)$
at the junction. This is the general relativistic generalization
of a thermal analog of the dc Josephson effect [18,19] according
to which the critical value of heat flow through a $SNS$ junction
is a nonmonotonic function of the magnetic flux $\Phi_b$.

Thus the gravitothermoelectric phenomena in superconductors
considered here allow us, in principle, to detect the general
relativistic effects. But we would like to emphasize that we only
concentrated on gravitothermoelectric phenomena in superconductors
of  type $I$. Nevertheless, recently several thermoelectric
effects were observed in high temperature superconducting
materials of type $II$ (see, for review, [5]). In this connection
further investigation is needed for taking into account
gravitational corrections for thermoelectric effects in type $II$
superconductors.

\section*{Acknowledgements}

The author acknowledges the financial support and hospitality at
the Abdus Salam International Centre for Theoretical Physics,
Trieste where this work has been completed. The research is also
supported in part by the UzFFR (project 01-06) and projects
F2.1.09, F2.2.06 and A13-226 of the UzCST.

\section*{References}
\begin{flushleft}

1. DeWitt, B. S. (1966). {\it Phys. Rev. Lett.
\bf 16}, 1092.\\
2. Papini, G. (1967). {\it Phys. Lett. \bf A24},
32.\\
3. Anandan, J. (1994). {\it Class. Quantum Grav. \bf 11},
A23; (1984). {\bf 1}, L51; (1984). {\it Phys. Lett. \bf A105},
280.\\
4. Jain, A.K., Lukens, J., and Tsai, J.S. (1987). {\it
Phys. Rev. Lett. \bf 58}, 1165.\\
5. Huebener, R.P. (1995). {\it Supercond. Sci. Technol. \bf 8},
189; Wang, Z.D., Wang, Q., and Fung, P.C.W. (1996). {\it
Supercond. Sci. Tecnol. \bf 9}, 333.\\
6. Geppert, U., and Wiebicke, H.-J. (1991). {\it Astron.
Astrophys. Supplement Series \bf 87}, 217.\\
7. Shapiro, S.L., and Teukolsky, S.A. (1983). {\it Black
Holes, White Dwarfs, and Neutron Stars} (New York: Wiley).\\
8.
Ahmedov, B.J. (1998). {\it Gravit. Cosmology \bf4}, 139.\\
9. Ahmedov, B.J. (1997). {\it Int. J. Mod. Phys. D \bf 6}, 341.\\
10. Van Harlingen, D.J. (1982). {\it Physica \bf 109\&110B},
1710.\\ 11. Van Harlingen, D.J., Heidel, D.F., and Garland, J.C.
(1980). {\it Phys. Rev. B \bf 21}, 1842.\\ 12. Lamb, F.K., (1991).
In {\it Frontiers of Stellar Evolution}, Lambert, D.L., ed. (San
Francisco: Astronomical Society of Pacific) 299. \\ 13. Aronov,
A.G., and Galperin, Yu.M. (1974). {\it JETP Lett. \bf 19}, 165.\\
14. Panaitov, G.I., Ryazanov, V.V., Ustinov, A.V., and Schmidt,
V.V. (1984). {\it Phys. Lett. A \bf 100}, 301.\\ 15. Ahmedov, B.J.
(1998). {\it Mod. Phys. Lett \bf B12}, 633.\\ 16. Clarke, J.
(1968). {\it Phys. Rev. Lett. \bf 21}, 1566.\\ 17. Michel, F.C.
(1991). {\it Theory of Neutron Star Magnetospheres} (Univ. Chicago
Press).\\ 18. Shmidt, V.V. (1981). {\it JETP Lett. \bf 33}, 98.\\
19. Ryazanov, V.V., and Schmidt, V.V. (1981). {\it Solid State
Comm. \bf 40}, 1055.\\
\end{flushleft}

\end{document}